\documentclass{article}
\usepackage[bookmarks=false, hidelinks]{hyperref}
\usepackage{relsize}
\usepackage{spconf,amsmath,graphicx}
\usepackage{color}
\usepackage{amssymb}

\title{LEARNING SOUND EVENT CLASSIFIERS FROM WEB AUDIO WITH NOISY LABELS}
\name{Eduardo Fonseca$^{\star}$\sthanks{This work is partially supported by the European Union's Horizon 2020 research and innovation programme under grant agreement No 688382 AudioCommons and a Google Faculty Research Award 2017.}, Manoj Plakal$^{\dagger}$, Daniel P. W. Ellis$^{\dagger}$, Frederic Font$^{\star}$, Xavier Favory$^{\star}$, Xavier Serra$^{\star}$}

\address{$^{\star}$Music Technology Group, Universitat Pompeu Fabra, Barcelona \{name.surname\}@upf.edu \\ 
$^{\dagger}$Google, Inc., New York, NY, USA \{plakal,dpwe\}@google.com}

\begin{document}
\ninept
\maketitle
\begin{abstract}
As sound event classification moves towards larger datasets, issues of label noise become inevitable.
Web sites can supply large volumes of user-contributed audio and metadata, but inferring labels from this metadata introduces errors due to unreliable inputs, and limitations in the mapping.
There is, however, little research into the impact of these errors.
To foster the investigation of label noise in sound event classification we present \textit{FSDnoisy18k}, a dataset containing 42.5 hours of audio across 20 sound classes, including a small amount of manually-labeled data and a larger quantity of real-world noisy data.
We characterize the label noise empirically, and provide a CNN baseline system.
Experiments suggest that training with large amounts of noisy data can outperform training with smaller amounts of carefully-labeled data.
We also show that noise-robust loss functions can be effective in improving performance in presence of corrupted labels.
\end{abstract}
\vspace{-2mm}
\section{Introduction}
\label{sec:intro}
\vspace{-1mm}
Data is essential to machine perception and, with the advent of deep learning, there is increasing demand for large-scale datasets to exploit the capacity of deep architectures.
In sound event classification, creating datasets for supervised learning 
typically consists of two stages:
\textit{i)} data acquisition (e.g., retrieving data from sites like Freesound or Youtube, or doing recordings) and \textit{ii)} data curation (organizing, cleaning and, most importantly, labeling the data).
Manual labeling is costly and is typically the limiting factor on audio datasets. 
Creators are often forced to compromise between dataset size and label quality.
Although some sound event datasets are exhaustively labeled, e.g., \cite{salamon2014dataset,piczak2015esc,foster2015chime} their size is limited (e.g., less than 9h of audio).
More recent datasets feature larger sizes, but their labeling is not as precise. 
For instance, AudioSet consists of 5000h labeled with 527 classes \cite{gemmeke2017audio}, 
but label error is estimated at above 50\% for $\approx$18\% of the classes.\footnote{See \url{https://research.google.com/audioset/dataset/index.html} for details on how the quality is estimated, accessed 22nd October 2018.}
FSDKaggle2018 \cite{fonseca2018general} is a dataset consisting of 18h of audio labeled with 41 classes, but only partially manually verified.
Hence we are witnessing a transition away from small and exhaustively labeled datasets in favour of larger datasets that inevitably include some amount of label noise.

Efficient creation of large-scale datasets from web audio requires minimizing curation effort.
We denote web audio as user-generated audio that is uploaded to online services such as Freesound and Youtube.
Labels can be inferred automatically from user-provided metadata, e.g., tags.
Such opportunistic labels support rapid collection of large amounts of data, but at the likely cost of a substantial level of label noise
arising from errors in 
the user-provided metadata or their transformation into labels.


In this context, label noise emerges as a pressing issue for the future of sound event classification.
The effects of label noise can include performance decrease, increased complexity of learned models, or changes in learning requirements \cite{frenay2014classification}.
To our knowledge, no previous audio dataset has specifically provided for the study of label noise.
Our first contribution is \textit{FSDnoisy18k}, an openly-available audio dataset that supports the investigation of real label noise, including an empirical characterization of the noise and a CNN baseline system. 
The dataset is singly-labeled and it consists of a small amount of clean data, and a much larger amount of noisy data containing a substantial amount of real-world label noise.

While the literature on label noise is extensive in computer vision, this topic has received little attention in sound event classification.
Some work focuses on self-training to learning from combinations of labeled and unlabeled data \cite{zhang2012semi,elizalde2017approach,han2016semi}, but the issue of label noise is not addressed per se.
In \cite{shah2018closer}, the effect of label noise on weakly supervised learning was analyzed by introducing noise to AudioSet. However, no measures to mitigate the effect of label noise were proposed.
In \cite{kumar2017audio}, classifiers are learnt from weakly labeled web data, and to improve performance an approach is proposed using a small amount of strongly labeled audio along with the web data.
In a recent audio tagging Kaggle competition using FSDKaggle2018 \cite{fonseca2018general}, a number of approaches were proposed to deal with the label noise present.
Some attempted to distinguish between the noisy and correct labels with the goal of selecting the latter, for instance, with self-training methods \cite{Dorfer2018,Nguyen2018}.
Others accepted the noisy labels but tried to mitigate their effect in the learning process.
Notably, one submission included a noise-robust loss function \cite{Jeong2018}, a technique requiring minimal intervention in the learning pipeline.
To motivate the usage of FSDnoisy18k as a resource for label noise research, our second contribution is an empirical evaluation of noise-robust loss functions using the proposed baseline system.
This is, to our knowledge, the first time that some of these loss functions have been used in sound classification.
This paper is organized as follows. 
In Section \ref{sec:dataset} we present FSDnoisy18k and characterize its label noise.
Section \ref{sec:baseline} describes a baseline system.
Section \ref{sec:method} introduces the noise-robust loss functions considered.
In Section \ref{sec:experi}, we discuss the results of a series of experiments.
Final remarks are given in Section~\ref{sec:conclusion}.
\vspace{-2mm}

\section{Dataset}
\label{sec:dataset}
\vspace{-1mm}
The source of audio content is Freesound---a sound sharing site hosting over 400,000 clips uploaded by its community of users, who additionally provide some basic metadata (e.g., tags, and title). 
More information about Freesound can be found in \cite{font2013freesound,Fonseca2017freesound}.
The 20 classes of FSDnoisy18k are drawn from the AudioSet Ontology: ``Acoustic guitar", ``Bass guitar", ``Clapping", ``Coin (dropping)", ``Crash cymbal", ``Dishes, pots, and pans", ``Engine", ``Fart", ``Fire", ``Fireworks", ``Glass", ``Hi-hat", ``Piano", ``Rain", ``Slam", ``Squeak", ``Tearing", ``Walk, footsteps", ``Wind", and ``Writing".
They are selected based on data availability as well as on their suitability to allow the study of label noise (see Section~\ref{ssec:dataset_noise} for some specific examples).
As a first step, we did a mapping of Freesound clips to the selected classes:
We assigned a number of Freesound tags to every class and, for each class, we selected the Freesound clips tagged with at least one of the tags. 
This process led to a number of automatically-generated \textit{candidate annotations} indicating the potential presence of a sound class in an audio clip. 
These annotations are at the clip level and hence are considered weak labels (although for some files the target signal fills the file entirely, which would be considered strongly-labeled).
Next, a small portion of the candidate annotations was human-validated. 
We used a validation task deployed in \emph{Freesound Annotator},\footnote{\url{https://annotator.freesound.org}\label{footnote_url_FSDs}} an online platform for the collaborative creation of open audio datasets \cite{Fonseca2017freesound}.
In this task, users verify the presence/absence of a candidate sound class in an audio clip with a \emph{rating} mechanism.
For every class, users are presented with a series of audio clips, and asked: \textit{Is $<$class$>$ present in the following sounds?} 
Users must select one of the responses: Present and Predominant (PP), Present but not Predominant (PNP), Not Present (NP) and Unsure (U).
The validation task is available online.\textsuperscript{\ref{footnote_url_FSDs}}

Audio clips that ended up with multiple labels had all but one label removed (in order to foster a type of label noise, see Section~\ref{ssec:dataset_noise}).
Next, we defined a \textit{clean} portion of the dataset consisting of correct and \textit{complete} labels, obtained by a second verification of the clips marked as PP. 
The remaining portion is referred to as the \textit{noisy} portion. 
The \textbf{clean portion} of the data consists of audio clips whose annotations are rated as PP (almost all with full inter-annotator agreement), meaning that the label is correct and, in most cases, there is no additional acoustic material other than the labeled class.
A few clips may contain some additional sound events, but they occur in the background and do not belong to any of the 20 target classes. 
This is more common for some classes that rarely occur alone, e.g., ``Fire", ``Glass" or ``Wind".
The \textbf{noisy portion} of the data consists of audio clips whose candidate annotations received no human validation, i.e., the only supervision comes from the user-provided tags. 
Hence, the noisy portion features a certain amount of label noise, which is characterized next.

\vspace{-2mm}
\subsection{Label Noise Characteristics}
\label{ssec:dataset_noise}

\begin{figure}[t]
     \vspace{-2mm}
    \centering
  \centerline{\includegraphics[width=0.99\columnwidth]{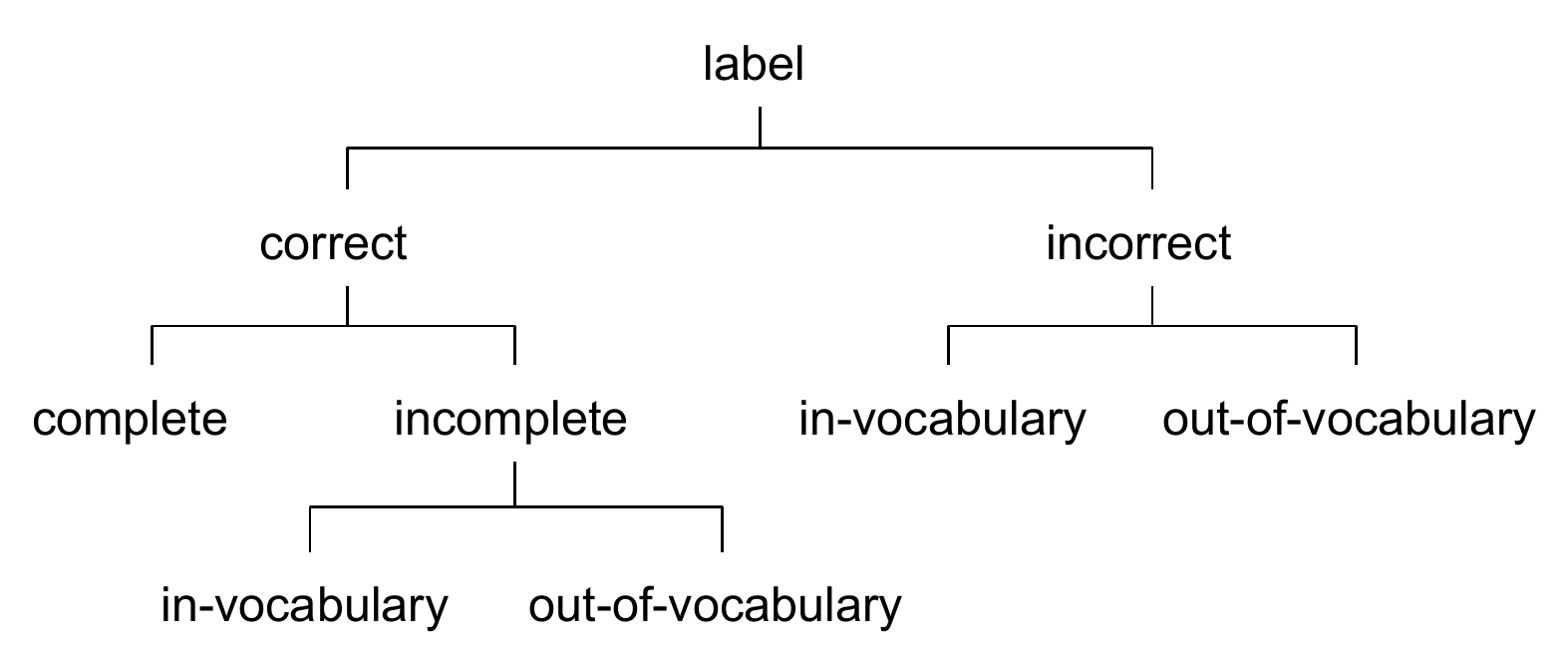}}
     \vspace{-2mm}
    \caption{Taxonomy of label noise based on the analysis of the noisy data in FSDnoisy18k.}
    \label{fig:taxo}
\end{figure}

The label noise literature typically deals with synthetic noise imposed on the data \cite{reed2014training,tanaka2018joint,zhang2018generalized}. 
Whereas synthetic label noise allows precise control of noise conditions, it may result in unrealistic conditions. 
FSDnoisy18k features real label noise that can be representative of audio data retrieved from the web, particularly from Freesound. 
In \cite{frenay2014classification}, a generic taxonomy of label noise from a statistical viewpoint is proposed, including models of label noise that differ in the dependencies among the agents involved.
In \cite{shah2018closer}, two types of label noise are proposed (a generic label corruption noise, and a label density noise) for multilabel data based on AudioSet.
We propose a taxonomy of label noise for singly-labeled data following an empirical approach.
The taxonomy is shown in Fig.~\ref{fig:taxo} and includes the noise types identified through manual inspection of a per-class, random, 15\% of the noisy data in FSDnoisy18k. 
Its concepts are explained next with the main use cases found in FSDnoisy18k. We distinguish between additional events that are already part of our target class set (``in-vocabulary'' or IV), or are not covered by those classes (``out-of-vocabulary'' or OOV). 
\cite{zhang2018generalized,wang2018iterative} use the terms \textit{closed-set} for IV, and \textit{open-set} for OOV.
Given an observed label that is incorrect or incomplete, the true or 
missing label can then be further classified as IV or OOV.

Some classes are prone to include incorrect labels when the clips are retrieved only on the basis of their existing user-provided tags, e.g., ``Bass guitar", ``Crash Cymbal", or ``Engine"; typically, the true label does not belong to the list of considered classes (\textit{incorrect/OOV}).
%
Other classes are prone to have audio clips with acoustic material that is additional to the provided (and correct) label, e.g., ``Rain", ``Fireworks" or ``Slam", and, again, the missing label usually does not belong to the list of considered classes (\textit{incomplete/OOV}).
Finally, a few classes are related to each other. 
It can happen that one class contains clips that actually belong to another class in the dataset, e.g. ``Wind" and ``Rain" (\textit{incorrect/IV}).
Alternatively, two sound classes can co-occur in an audio clip, e.g. ``Slam" and ``Squeak", despite only a single label is available (\textit{incomplete/IV}).
For completeness, correct and complete labels mean no label noise, i.e., clean data.

In addition to the aforementioned noise types, two more types arise in the context of web audio and Freesound in particular.
First, determining whether a sound class is present in an audio clip can be subjective, even for an expert. 
This happens with human imitations or heavily processed sounds (e.g., with sound effects).
We refer to these clips as \textit{ambiguous} as it is unclear whether the label is correct or not.
The second noise type relates to \textit{i)} the variable clip lengths and \textit{ii)} the weak nature of the clip-level labels.
A naive but common way of processing variable-length clips is to split them into several fixed-length patches, each inheriting the clip-level label (called \textit{false strong labeling} in \cite{morfi2018data}).
This can generate false positives if the label is not active in a given patch. 
This type of label noise is conceptually similar to the label density noise of \cite{shah2018closer}.

\begin{table}[t]
\vspace{-4mm}
\caption{Distribution of label noise types in a random 15\% of the noisy data of FSDnoisy18k.}
\centering
\begin{tabular}{lc|lc}
\textbf{Label noise type}   & \textbf{Amount}  &\textbf{Label noise type} & \textbf{Amount} \\
\hline
Overall      			    & 60\%           & Incorrect/IV        & 6\%        \\
Incorrect/OOV 			& 38\%           &Incomplete/IV        & 5\%        \\
Incomplete/OOV         & 10\%           & Ambiguous labels            & 1\%        \\
\end{tabular}
\vspace{-5mm}
\label{tab:noise_types}
\end{table}

The analysis of the noisy data revealed that roughly 40\% of the analyzed labels are correct and complete, whereas 60\% of the labels show some type of label noise, whose distribution is listed in Table~\ref{tab:noise_types}.
The most frequent types of label noise correspond to out-of-vocabulary (OOV) problems, either in the form of incorrect labels (that generate false positives) or incomplete labels (which generate false negatives). 
Furthermore, we have observed that a few clips within the incorrect/OOV category are incorrectly labeled according to the semantic meaning of the class, and yet they are relatively similar (in terms of their acoustics) to the true label.
For example, in ``Clapping" there is a certain amount of applause sounds and claps generated by drum machines.
We estimate that $\approx$10\% of the clips analyzed shows this phenomenon, although it is highly subjective.
This $\approx$10\% is included in the 38\% of incorrect/OOV labels.
The label density noise is only relevant in few classes, especially ``Slam", and to a lesser extent ``Fireworks" and ``Fire".
This type of noise was quantified by counting the audio clips that present at least one segment of 2s (or more) where the observed label is not present (2s is the patch length used in the baseline system, see Section~\ref{sec:baseline}).
The degree of total label noise per-class ranges from 20\% to 80\% roughly. 
A per-class description of the label noise similar to that of Table~\ref{tab:noise_types} is available at the dataset companion site in order to facilitate per-class analysis.\footnote{\url{http://www.eduardofonseca.net/FSDnoisy18k/}\label{companion}} 

\vspace{-2mm}
\subsection{Dataset Characteristics}
\label{ssec:dataset_charac}

\begin{figure}[t]
    \vspace{-1mm}
    \centering
  \centerline{\includegraphics[width=\columnwidth]{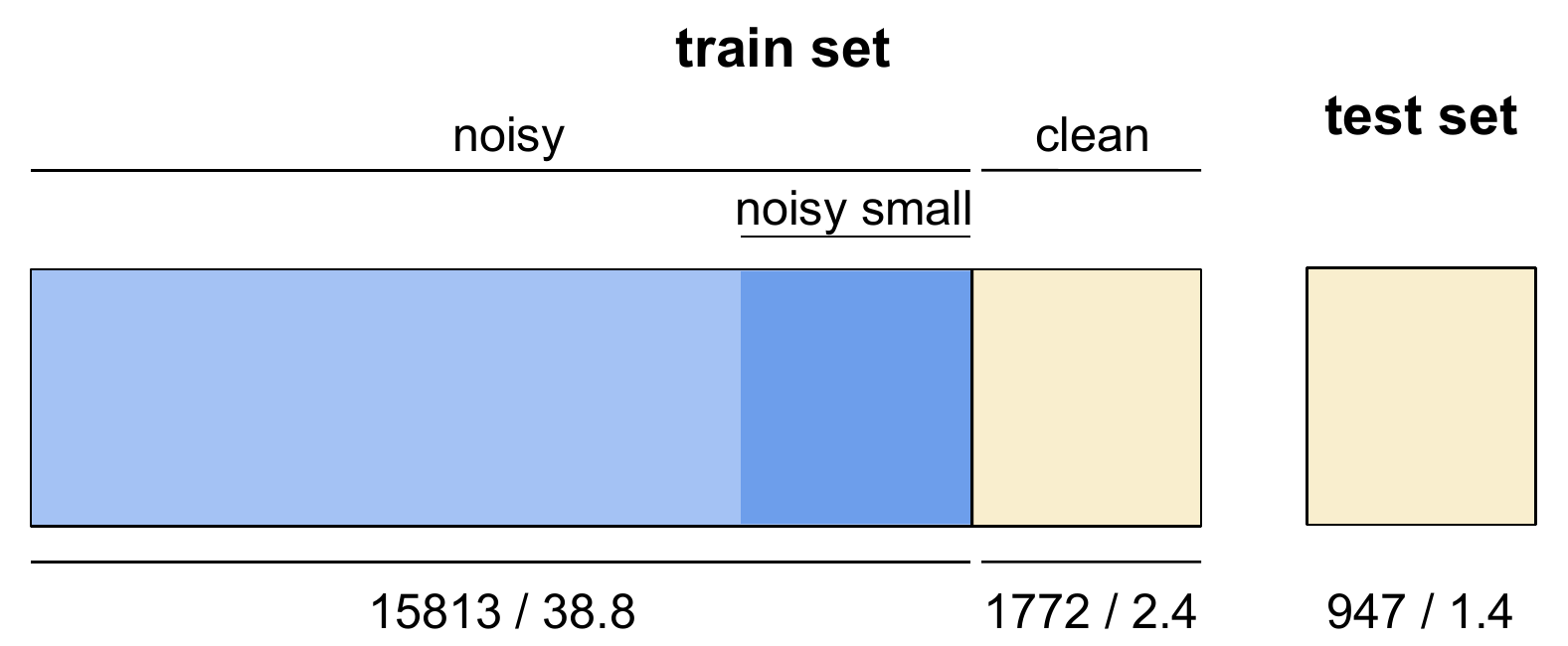}}
     \vspace{-2mm}
    \caption{Data split in FSDnoisy18k, including number of clips / duration in hours. Blue = noisy data. Yellow = clean data.}
    \label{fig:split}
\vspace{-5mm}
\end{figure}

FSDnoisy18k contains 18,532 mono audio clips (42.5h) unequally distributed in the 20 aforementioned classes drawn from the AudioSet Ontology.
The audio clips are of variable length ranging from 300ms to 30s, and each clip has a single ground truth label (singly-labeled data). 
The dataset is split into a test set and a train set as seen in Fig.~\ref{fig:split}.
The test set is drawn entirely from the clean portion, while the remainder of data forms the train set.
The \textbf{train set} is composed of 17,585 clips (41.1h) unequally distributed among the 20 classes.
It features a clean subset and a noisy subset.
In terms of number of clips their proportion is $\approx$10\%/90\%, whereas in terms of duration the proportion is slightly more extreme ($\approx$6\%/94\%).
The per-class percentage of clean data within the train set is also imbalanced, ranging from 6.1\% to 22.4\%. 
The number of audio clips per class ranges from 51 to 170, and from 250 to 1000 in the clean and noisy subsets, respectively.
Further, a \textit{noisy\_small} subset is defined (dark blue box in Fig.~\ref{fig:split}), which includes an amount of (noisy) data comparable (in terms of duration) to that of the clean subset.
The \textbf{test set} is composed of 947 clips (1.4h) that belong to the clean portion of the data.
Its class distribution is similar to that of the clean subset of the train set. 
The number of per-class audio clips in the test set ranges from 30 to 72.
The test set enables a multi-class classification problem.
The dataset is openly available from its companion site,\textsuperscript{\ref{companion}} along with the proposed data splits for reproducibility and a more detailed dataset description.
FSDnoisy18k is an expandable dataset that features a per-class varying degree of types and amount of label noise.
The dataset allows investigation of label noise as well as other approaches, from semi-supervised learning, e.g., self-training \cite{elizalde2017approach} to learning with minimal supervision \cite{veit2017learning}.
\vspace{-1mm}

\section{Baseline System}
\label{sec:baseline}
\vspace{-1mm}
Incoming audio is transformed to 96-band, log-mel spectrogram as input representation.
To deal with the variable-length clips, we use time-frequency patches of 2s; 
shorter clips are replicated while longer clips are trimmed in several patches inheriting the clip-level label.
The model used is a CNN (3 conv + 1 dense) following that of \cite{salamon2017deep} with two main changes.
First, we include Batch Normalization (BN) \cite{ioffe2015batch} between each convolutional layer and ReLU non-linearity.
Second, we use \textit{pre-activation}, a technique initially devised in deep residual networks \cite{he2016identity} which essentially consists of applying BN and ReLU  as pre-activation before each convolutional layer.
It was proved beneficial for acoustic scene classification in \cite{fonseca2018simple}, as well as in preliminary experiments with FSDnoisy18k. 
The model has $\approx$0.5M weights.
The loss function is categorical cross-entropy (CCE), the batch size is 64, and we use Adam optimizer \cite{kingma2014adam} with initial learning rate of 0.001, which is halved whenever the validation accuracy plateaus for 5 epochs.
Earlystopping is adopted with a patience of 15 epochs on the validation accuracy.
To this end, a 15\% validation set is split randomly from the training data of every class.
The system is implemented in Keras and TensorFlow.
The prediction for every clip is obtained by computing predictions at the patch level, and aggregating them with geometric mean to produce a clip-level prediction.
The goal of the baseline is to give a sense of the classification accuracy that a well-known architecture can attain and not to maximize the performance. 
Extensive hyper-parameter tuning or additional model exploration was not conducted.
The code and a more detailed description of the baseline are available at \footnote{\url{https://github.com/edufonseca/icassp19}}.


\section{Noise-Robust Loss Functions}
\label{sec:method}
The training of a deep network is based on updating the network weights to minimize a loss function that expresses the divergence between the network predictions and the ground truth labels.
If the ground truth labels are corrupted, the weights' update can be suboptimal thus hindering model convergence.
In these cases, loss functions that are robust against label noise can be helpful.
Next, we briefly describe the noise-robust loss functions evaluated and their underlying principles.
All of them are modifications of the CCE loss commonly-used for multi-class classification.
The reader is referred to the original papers for further details.
The CCE loss is given by (\ref{eqn:cce}), where $y_{k}$ is the $k$'th element of the target label (a one-hot encoded vector), $\hat{y}_{k}$ is the $k$'th element of the network predictions (the predicted class probabilities), and $K$ is the number of classes:
\vspace{-1mm}
\begin{equation}
\vspace{-1mm}
\label{eqn:cce}
L_{cce}=-\sum_{k=1}^Ky_{k}\log(\hat{y}_{k}).
\end{equation}

The $L_{soft}$ loss function dynamically updates the target labels based on the current state of the model. 
More specifically, the updated target label is a convex combination of the current model's prediction and the (potentially noisy) target label.
The idea is to pay less attention to the noisy labels, in favour of the model predictions, which are more reliable as the learning progresses.
This approach is referred to as soft bootstrapping \cite{reed2014training} and is expressed by (\ref{eqn:reed}):
\vspace{-1mm}
\begin{equation}
\vspace{-1mm}
\label{eqn:reed}
L_{soft}=-\sum_{k=1}^K[\beta y_{k}+(1-\beta)\hat{y}_{k}]\log(\hat{y}_{k}), \quad \beta \in [0,1].
\end{equation}

The $L_{q}$ loss is a generalization of CCE and mean absolute error (MAE) proposed in \cite{zhang2018generalized}.
In CCE, the predictions that differ more from the target labels are also weighed more for the gradient update. 
This is beneficial when dealing with clean data but it can be undesirable in the case of noisy labels.
On the contrary, MAE weighs all the predictions equally (which, in theory, makes it robust against corrupted labels \cite{ghosh2017robust}). 
However, in preliminary experiments we obtained poor performance using MAE with FSDnoisy18k, in accordance with findings reported in \cite{zhang2018generalized} with other datasets.
The $L_{q}$ loss aims to take advantage of the benefits of both CCE and MAE, and is given by (\ref{eqn:zhang}):
\vspace{-1mm}
\begin{equation}
\vspace{-1mm}
\label{eqn:zhang}
L_{q}=\frac{1-(\sum_{k=1}^K y_k \hat{y}_k)^q}{q}, \quad q \in [0,1].
\end{equation}

The last approach consists of first computing the CCE loss function, and then applying heuristics to discard loss values that may come from data points with corrupted labels.
Intuitively, when labels are corrupted, model predictions are likely to be less congruent with the noisy target labels, yielding artificially high losses.
By discarding the latter, we prevent the data points in the minibatch that presumably feature corrupted labels from contributing to the total loss.
This can be understood as a loss masking approach.
First, we compute the CCE loss for every data point in the minibatch, $\boldsymbol{L_{cce}} \in \mathbb{R}^{64\times1}$.
Then, we define a threshold $t$, such that elements in $\boldsymbol{L_{cce}}$ greater than $t$ are discarded for the computation of the total loss.
We experiment with two thresholds: $t_m = m\cdot max(\boldsymbol{L_{cce}})$ proposed in \cite{Jeong2018}, and $t_l = median(\boldsymbol{L_{cce}}) + l\cdot \sigma(\boldsymbol{L_{cce}})$, where $m \in [0,1]$, $l \in [0,\infty)$ and $\sigma$ is standard deviation.
These thresholds correspond to the rows labeled $L_{m}$ and $L_{l}$, respectively, in Table~\ref{tab:results}.
\vspace{-1mm}


\section{Experiments and Discussion}
\label{sec:experi}
\vspace{-1mm}
We present the experiments carried out with the baseline system and the proposed noise-robust loss functions, evaluated by replacing the CCE loss by each one of them in the baseline system.
To ensure a fair comparison, the learning setup of Section~\ref{sec:baseline} is always kept.
\vspace{-2mm}
\subsection{Baseline System}
\label{ssec:experi_baseline}

The results for different subsets of training are listed in the first row of Table~\ref{tab:results}.
From right to left, it can be seen that using the \textit{clean} subset leads to an accuracy increase\footnote{Performance differences are expressed in terms of absolute accuracy.} of 15.8\% with respect to using the \textit{noisy\_small} subset (consisting of roughly the same amount of data).
However, curating the clean subset requires significant effort.
Training with the \textit{noisy} subset provides a boost of 6.3\% over the performance obtained with the clean subset (despite the considerable amount of label noise present).
Nevertheless, this improvement comes at the expense of using data that is an order of magnitude greater (see Section~\ref{ssec:dataset_charac}). 
Finally, using the entire train set, that is, adding a small amount of manually-curated data to the noisy subset, increases the accuracy by 5.1\%.
The results suggest that large amounts of Freesound audio with the level of supervision provided by the user-generated tags can be a feasible option for training sound event recognizers. 
This can be useful in case of no labeling budget, as long as the computational resources can be accommodated.
If only limited budget is available, curating a small portion of data to be combined with large amounts of noisy data yields top performance.

\begin{table}[t!]
\vspace{-4mm}
\caption{Average classification accuracy (\%) and 95\% confidence interval (7 runs) obtained by several approaches using different subsets of FSDnoisy18k for training (see Fig.~\ref{fig:split}); \textit{all} = entire train set.}
\centering
\begin{tabular}{@{}l@{}|cccc@{}}
\textbf{Approach} 	& \textbf{all}   & \textbf{noisy}  & \textbf{noisy\_small}   & \textbf{clean} \\
\hline
\textbf{baseline}	& 71.6$\pm0.4$	    & 66.5$\pm0.6$        	& 44.4$\pm1.1$		& 60.2$\pm0.5$\\
\hline
$L_{soft},\beta=0.3 $  	& 73.1$\pm0.6$ 		    & 66.8$\pm0.6$ 	        & \textbf{46.0$\pm0.9$} & -- \\
$L_{soft},\beta=0.7$ 	& 72.6$\pm0.6$	        & 67.6$\pm0.7$ 	        & 44.6$\pm1.0$ 		& -- \\
\hline

$L_{q}, q=0.5$     & 73.4$\pm0.8$ 		    & \textbf{68.4$\pm0.5$}     & 45.0$\pm1.0$ 		& -- \\
$L_{q}, q=0.7$ 	& \textbf{74.3$\pm0.7$}	    & 66.7$\pm1.2$              & 43.2$\pm1.2$ 		& -- \\
\hline

$L_{m}, m=0.5$		& 71.5$\pm0.5$	        & 67.7$\pm0.9$          & 45.4$\pm1.1$  & -- \\
$L_{m}, m=0.6$     & 72.2$\pm0.7$	        & 66.9$\pm0.8$          & 45.7$\pm1.2$ 		& -- \\
\hline

$L_{l}, l=1.9$     & 71.8$\pm1.0$	    & 67.2$\pm0.7$         	& 44.6$\pm1.0$ 		& -- \\
$L_{l}, l=2.0$     & 71.5$\pm0.7$	    & 67.6$\pm0.8$          & 44.5$\pm1.0$ 		& -- \\
\end{tabular}
\label{tab:results}
\vspace{-5mm}
\end{table}

\vspace{-2mm}
\subsection{Noise-Robust Loss Functions}
\label{ssec:experi_loss}
Classification accuracy results for different subsets of training data and loss functions are listed in Table~\ref{tab:results}.
We show results after fine-tuning the hyper-parameter of every loss function.
When training with the entire train set or the noisy subset, the top-performing loss function is consistently $L_{q}$, followed by $L_{soft}$, and finally followed by the heuristics-based approaches (where $L_{m}$ shows a modest improvement over $L_{l}$).
This means that $L_{soft}$, and especially $L_{q}$, (originally proposed for image recognition) also work well for sound classification tasks.
More specifically, $L_{q}$ provides an accuracy increase over the baseline of 2.7\% and 1.9\% for the entire train set and noisy subset, respectively.
The results confirm the insights in \cite{zhang2018generalized}, where it is shown that $L_{q}$ works well with both OOV and IV noisy labels, which is the case of FSDnoisy18k.

When training with the entire train set, the noise-robust approaches of Section~\ref{sec:method} are applied selectively based on data origin, i.e., they are applied only to the data coming from the noisy subset, whereas for the clean subset the regular CCE loss is adopted.
Specifically, this means: \textit{i)} in $L_{soft}$ the target labels are updated only for data points coming from the noisy subset; 
\textit{ii)} when testing $L_{q}$, only data points from the noisy subset contribute with $L_{q}$ to the total loss;
\textit{iii)} in the heuristics-based approaches, the computation is as described in Section~\ref{sec:method} except that only data points from the noisy subset are susceptible to be discarded.
For $L_{soft}$, and especially $L_{q}$, this selective procedure leads to slightly better performance, in contrast to the naive way of mixing all the data and applying the noise-robust approaches indiscriminately.
This suggests that $L_{q}$ is more effective when a greater amount of label noise is present.


It is interesting to compare \textit{i)} the accuracy boost obtained from adding manually-curated data to the noisy subset, versus \textit{ii)} the boost resulting from using the noisy subset with the top-performing loss function.
The baseline classification accuracy when training with the noisy subset is 66.5\%.
If we add a small amount of curated data, we obtain a 5.1\% boost (see \textit{all} column).
Conversely, if we leverage the top performing $L_{q}$ we obtain an increase of 1.9\% (i.e., $\approx$37\% of the boost by manual curation).
Note that the manual curation requires a significant effort, while the latter approach requires very little engineering effort and adds minimal computational cost.
Combining both approaches yields top performance.
\vspace{-1mm}


%

\section{Conclusion}
\label{sec:conclusion}
\vspace{-1mm}
We have presented FSDnoisy18k, an openly-available dataset that facilitates the investigation of label noise in sound event classification.
The dataset is singly-labeled and consists of a small amount of manually-labelled data and a large amount of noisy data, featuring a per-class varying degree of types and amount of real label noise.
An empirical characterization of the dataset reveals that the noisy data presents $\approx$60\% of label noise, most of which corresponds to OOV noisy labels.
Experiments with a CNN baseline system suggest that large amounts of Freesound audio with the level of supervision provided by the user-generated tags can be a feasible option for training sound event recognizers.
In addition, the evaluation of four noise-robust loss functions shows that some of them, originally proposed for image recognition, are an efficient way to significantly improve performance in presence of corrupted labels, while requiring minimal engineering effort.
FSDnoisy18k opens the door to the evaluation of a variety of measures against label noise, as well as to a number of semi-supervised learning approaches.
It may also be interesting to evaluate the proposed noise-robust loss functions on larger amounts of noisy data and with models of larger capacity.




\vfill\pagebreak
\section{Acknowledgments}
We thank Zhilu Zhang of Cornell University for his assistance with the $L_{q}$ loss, and we thank everyone who contributed to FSDnoisy18k with annotations, especially Mercedes Collado.
Eduardo Fonseca is also grateful for the GPU donated by NVidia.

\bibliographystyle{IEEEbib}
\bibliography{strings,refs}

\end{document}